\documentclass[11pt,aps,showpacs,showkeys]{revtex4}
\usepackage{amsmath,amsthm,amssymb,epsfig,alltt}

\begin{document}

\def\a{\alpha}
\def\b{\beta}
\def\d{{\delta}}
\def\l{\lambda}
\def\e{\epsilon}
\def\p{\partial}
\def\m{\mu}
\def\n{\nu}
\def\t{\tau}
\def\th{\theta}
\def\s{\sigma}
\def\g{\gamma}
\def\o{\omega}
\def\r{\rho}
\def\z{\zeta}
\def\D{\Delta}
\def\half{\frac{1}{2}}
\def\hatt{{\hat t}}
\def\hatx{{\hat x}}
\def\hatp{{\hat p}}
\def\hatX{{\hat X}}
\def\hatY{{\hat Y}}
\def\hatP{{\hat P}}
\def\haty{{\hat y}}
\def\whatX{{\widehat{X}}}
\def\whata{{\widehat{\alpha}}}
\def\whatb{{\widehat{\beta}}}
\def\whatV{{\widehat{V}}}
\def\hatth{{\hat \theta}}
\def\hatta{{\hat \tau}}
\def\hatrh{{\hat \rho}}
\def\hatva{{\hat \varphi}}
\def\barx{{\bar x}}
\def\bary{{\bar y}}
\def\barz{{\bar z}}
\def\baro{{\bar \omega}}
\def\barpsi{{\bar \psi}}
\def\sp{\sigma^\prime}
\def\nn{\nonumber}
\def\cb{{\cal B}}
\def\2pap{2\pi\alpha^\prime}
\def\pap{\pi\alpha^\prime}
\def\wideA{\widehat{A}}
\def\wideF{\widehat{F}}
\def\beq{\begin{eqnarray}}
 \def\eeq{\end{eqnarray}}
 \def\4pap{4\pi\a^\prime}
 \def\op{\omega^\prime}
 \def\xp{{x^\prime}}
 \def\sp{{\s^\prime}}
 \def\ap{\a^\prime}
 \def\tp{{\t^\prime}}
 \def\zp{{z^\prime}}
 \def\rp{{\rho^\prime}}
  \def\spp{\s^{\prime\prime}}
 \def\xpp{x^{\prime\prime}}
 \def\xppp{x^{\prime\prime\prime}}
 \def\barxp{{\bar x}^\prime}
 \def\barzp{{\bar z}^\prime}
 \def\barxpp{{\bar x}^{\prime\prime}}
 \def\barxppp{{\bar x}^{\prime\prime\prime}}
 \def\zetap{{\zeta^\prime}}
 \def\barchi{{\bar \chi}}
 \def\baro{{\bar \omega}}
 \def\bpsi{{\bar \psi}}
 \def\barg{{\bar g}}
 \def\barz{{\bar z}}
 \def\bareta{{\bar \eta}}
 \def\ta{{\tilde \a}}
 \def\tb{{\tilde \b}}
 \def\tc{{\tilde c}}
 \def\tz{{\tilde z}}
 \def\tJ{{\tilde J}}
 \def\tpsi{\tilde{\psi}}
 \def\tal{{\tilde \alpha}}
 \def\tbe{{\tilde \beta}}
 \def\tga{{\tilde \gamma}}
 \def\tchi{{\tilde{\chi}}}
 \def\barth{{\bar \theta}}
 \def\bareta{{\bar \eta}}
 \def\barom{{\bar \omega}}
 \def\bole{{\boldsymbol \epsilon}}
 \def\bolth{{\boldsymbol \theta}}
 \def\bomega{{\boldsymbol \omega}}
 \def\bolmu{{\boldsymbol \mu}}
 \def\bolal{{\boldsymbol \alpha}}
 \def\bolbe{{\boldsymbol \beta}}
 \def\bolL{{\boldsymbol  L}}
 \def\bolX{{\boldsymbol X}}
 \def\bbk{{\boldsymbol k}}
 \def\boln{{\boldsymbol n}}
 \def\bols{{\boldsymbol s}}
 \def\bolS{{\boldsymbol S}}
 \def\bola{{\boldsymbol a}}
 \def\bolA{{\boldsymbol A}}
 \def\bolvarphi{{\boldsymbol \varphi}}
 \def\boleta{{\boldsymbol \eta}}
 \def\bolchi{{\boldsymbol \chi}}
 \def\bolJ{{\boldsymbol J}}
 \def\tr{{\rm tr}}
 \def\bbP{{\mathbb P}}
 \def\bbp{{\boldsymbol p}}
 \def\mathP{{\mathbb P}}

\setcounter{page}{1}
\title[]{Witten's Cubic Open String Field Theory on multiple $Dp$-branes}

\author{Taejin Lee}
\email{taejin@kangwon.ac.kr}
\affiliation{Department of Physics, Kangwon National University, Chuncheon 24341 Korea}

\date{\today }

\begin{abstract}
We study Witten's cubic open string field theory on multiple $Dp$-branes. On multiple $Dp$-branes the string fields carry $U(N)$ group indices. Mapping the string world-sheet onto the upper half complex plane and evaluating the Polyakov string path integral of Witten's cubic open string,
we obtain a Fock space representation of one, two and three-string vertex operators. In the low energy region, the system is described by massless gauge fields and massless scalar fields, carrying $U(N)$ group indices. The two-string vertex induces mass terms for the gauge fields and the scalar fields. The cubic interaction reduces correctly to the cubic term of the non-Abelian gauge fields, cubic interaction terms of the gauge fields, and the scalar fields on $(p+1)$ dimensional space. Thus, Witten's string field theory may be a useful tool to study the quantum dynamics of open strings on multiple $Dp$-branes.   
\end{abstract}




\maketitle

\setcounter{footnote}{0}

\newpage
\tableofcontents
\newpage

\section{Introduction}

The covariant open string field theory \cite{Witten1986,Witten92p,Banks1986} which contains only the cubic interaction, has been proven to be a useful tool for exploring string theory corrections \cite{Lee2016i,Lee2017d}, entanglement entropy in string theory \cite{TLeeEnt2018}, and the off-shell dynamics of string theory such as tachyon condensation \cite{sen1998,SenZ2000,Tlee:01nc,Tlee:01os}. The cubic string field theory of Witten on multiple $Dp$-branes is also expected to be useful for studying the quantum dynamics of string theory in a systematic manner. 
Therefore, it is important to develop cubic string field theory as a practical tool to study 
various subjects in theoretical physics. String field theories are usually constructed on the 
configuration space, with overlapping conditions. However, in order to apply the string field theory to particle physics, we need a Fock space representation of the theory. It is not an easy task to convert 
the configuration space representation of string field theory into the Fock space representation, as 
noted earlier in a work on the light-cone string field theory \cite{Kaku1974a,Kaku1974b}. In the light-cone string field theory, the correct Fock space representation of the string field theory is obtained by using the 
Neumann functions on the world sheets of strings, which are mapped onto the upper half-plane \cite{Mandelstam1973,Mandelstam1974,cremmer1975}.
The purpose of this work is to extend the previous work on the covariant string field theory on multiple 
$Dp$-branes in the proper-time gauge \cite{TLee2017cov} to Witten's cubic open string field theory on multiple $Dp$-branes. Our main task is to construct the three-string vertex operator. 

In the absence of $Dp$-branes, the three-string vertex operator for Witten's cubic string field theory has been constructed in \cite{Cremmer86,Samuel86,Ohta1986,Grossjevicki87a,Grossjevicki87b}. The Witten's cubic string field theory has been discussed also in the presence of the $Dp$-branes in \cite{Harvey2000,Matsuo2001,Rastelli2001,Michishita2001,Imamura2002} before. But these previous works mostly focused on the tachyonic vacuum on a $D25$-brane. Refs. \cite{Moeller2000,Moeller2000b,Taylor2000,Gaiotto2003,Coletti05} discussed the effective field theory action of $Dp$-branes, using numerical method, called the level truncation. Relationships between the string field theory and matrix models are studied in \cite{Ikehara1995,Fukuma1998,Ennyu2003,Takayanagi2005,Zeze2016} and scattering of strings from D-branes has been explored in detail in \cite{Hashimoto1997a,Chan2007,J.-C. Lee2007a,J.-C. Lee2008,J.-C. Lee2011}.

Two approaches to the covariant string theory exist: the first quantized theory,
which is based on the Polyakov string path integral \cite{Polyakov1981}, and the second quantized theory, which is formulated as the cubic string field theory. We shall establish these two approaches
for string theory on multiple $Dp$-branes 
by explicitly constructing Fock space representations of string vertex operators. The cubic string field theory 
provides a unique way to parameterize the world sheet diagram in terms of local flat coordinate patches 
which describes the free string propagation. The world sheet of three strings in the 
cubic string field theory forms a conical surface with an excessive angle $\pi$. It can be mapped onto 
a unit disk on which the Neumann boundary condition is imposed. Then, the unit disk is mapped again onto the upper half-plane. This procedure defines the Schwarz-Christoffel 
transformation from the world sheet onto the upper half-plane and fixes the Green's function on the world sheet of three strings. 
Using the Green's function on the world sheet, we can evaluate the Polyakov string path integral which 
is subject to the temporal boundary condition, fixed by the momenta of external string states. The 
Fock space representation of the three-string vertex is obtained by rewriting the Polyakov string path integral in terms of the oscillator operators. As a result, we obtain explicit expressions of the Neumann functions of the corresponding Vertex operators, {\it i.e.}, the Fock space representation of the vertex operators. 

After constructing the one, two, and three-string vertex operators, we calculate scattering amplitudes in the 
low energy region, where only massless gauge fields and massless scalar fields are excited. The one-string vertex operator corresponds to insertion of field operators with zero momenta, and the two-string vertex operator yields mass terms of the component fields as expected. From the three-string vertex operator we obtain the cubic term in the covariant non-Abelian gauge field action and the covariant coupling of the scalar field and the gauge field.
If we choose higher spin excited states, we would obtain their consistent couplings from the vertex operators. We conclude with a brief summary and discussion on possible extensions and applications of the present work in various areas in string theory and related subjects.

\section{Witten's Cubic Open String Fields on $Dp$-branes}

\begin{figure}[htbp]
\begin {center}
\epsfxsize=0.6\hsize

\epsfbox{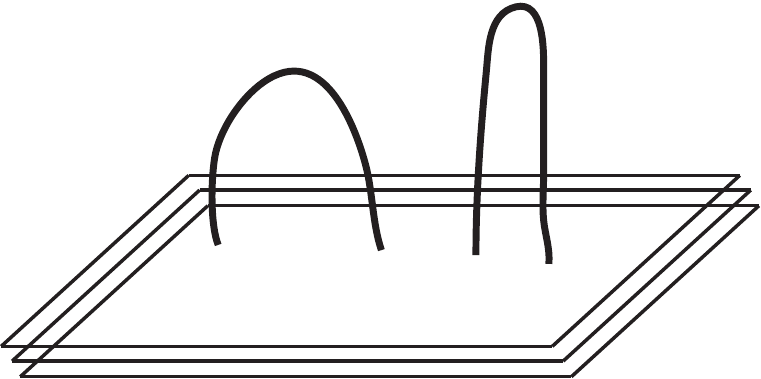}
\end {center}
\caption {\label{Dpbrane} Open strings on $Dp$-branes.}
\end{figure}

\begin{figure}[htbp]
\begin {center}
\epsfxsize=0.8\hsize

\epsfbox{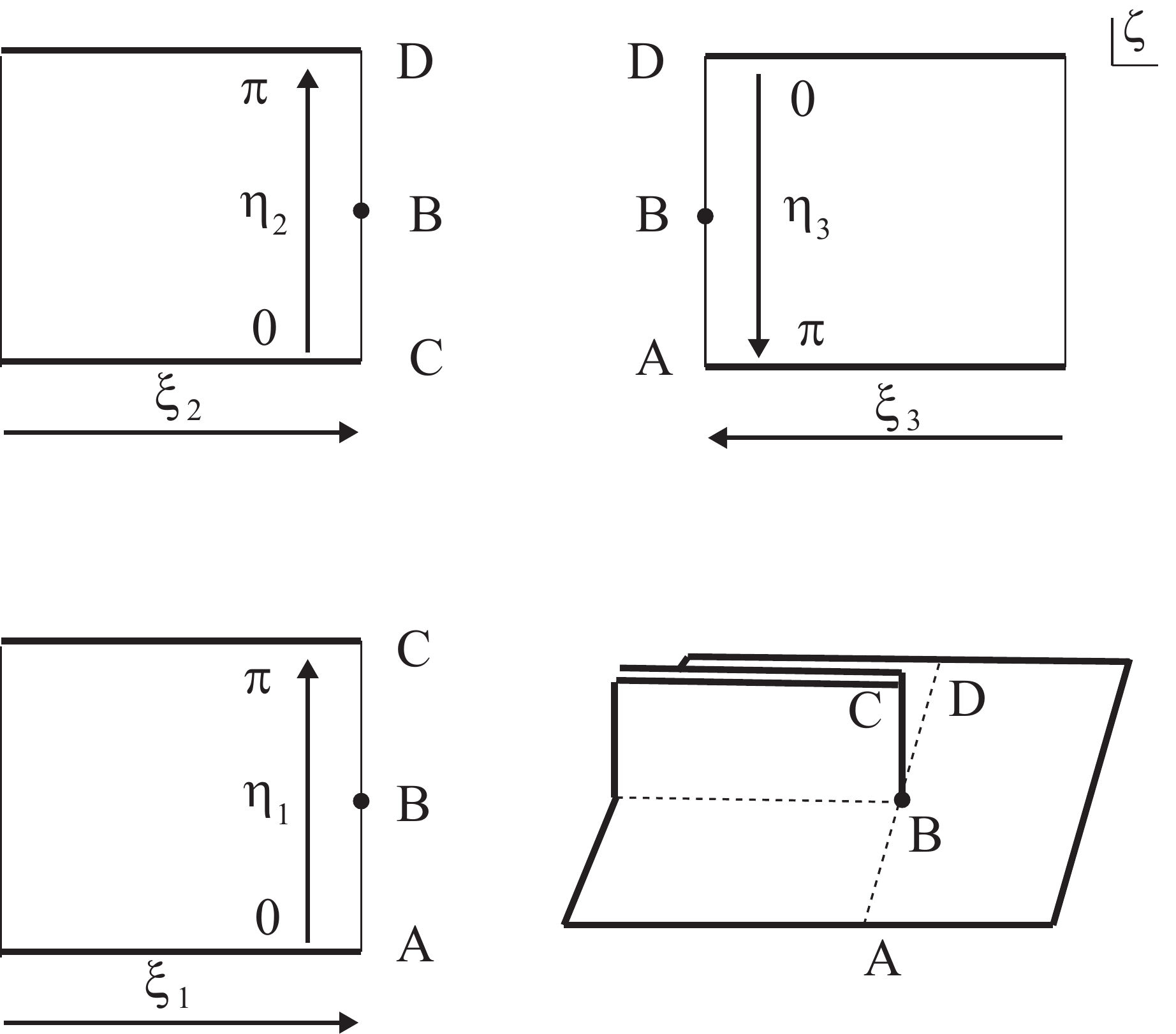}
\end {center}
\caption {\label{3patches1} The world-sheet of three-open-string interaction.}
\end{figure}

Witten's cubic open 
string field theory is described by a BRST invariant action, which has only a cubic interaction term
\beq
S_{\rm open} &=& \int \text{tr} \left( \Psi * Q \Psi + \frac{2g}{3} \Psi * \Psi * \Psi \right),
\eeq
where the star product between the string field operators is defined as
\beq
\left(\Psi_1 * \Psi_2\right) [X(\s)] &=& \int \prod_{\frac{\pi}{2} \le \s \le \pi} DX^{(1)}(\s) \prod_{0 \le \s \le \frac{\pi}{2}} DX^{(2)}(\s)  \nn\\
&& 
\prod_{\frac{\pi}{2} \le \s \le \pi} \d \left[X^{(1)}(\s) - X^{(2)}(\pi -\s) \right] \Psi[X^{(1)}(\s)] \Psi[X^{(2)}(\s)]. \label{star}
\eeq 
The star product is associative and the string field action is invariant under the BRST gauge transformation
\beq \label{BRST}
\d \Psi = Q * \e + \Psi * \e - \e * \Psi .  
\eeq 
On $Dp$-branes, in terms of the normal modes, 
the open string coordinates $X^I$, $I = 0, \dots, d$, are expanded as
\beq
X^\m (\s) &=& x^\m + 2 \sum_{n=1}^\infty \frac{1}{\sqrt{n}} x^\m_n \cos \left(n \s\right), ~~ \m = 0, 1, \dots, p, \nn\\
X^i(\s) &=& 2 \sum_{n=1} \frac{1}{\sqrt{n}} \sin (n\s), ~~ i=p+1, \dots,d.
\eeq 
The string coordinates tangential to the $Dp$-branes satisfy the Neumann boundary condition, and the 
string coordinates orthogonal to the $Dp$-branes satisfy the Dirichlet boundary condition. (See Fig. \ref{Dpbrane})
The string field $\Psi$ may carry the group indices:
\beq
\Psi[X]  =\frac{1}{\sqrt{2}} \Psi^0 [X] + \Psi^a[X] T^a,  ~~~ a =1, \dots, N^2-1 ,
\eeq
where $\Psi^0$ is the $U(1)$ component and $\Psi^a$, $a =1, \dots, N^2-1$ are the $SU(N)$ components. 
If we introduced three local coordinate patches, which describe propagation of three open strings,
we might have depicted the string world-sheet of three-open-string interaction by Fig. \ref{3patches1}.

\begin{figure}[htbp]
\begin {center}
\epsfxsize=0.7\hsize

\epsfbox{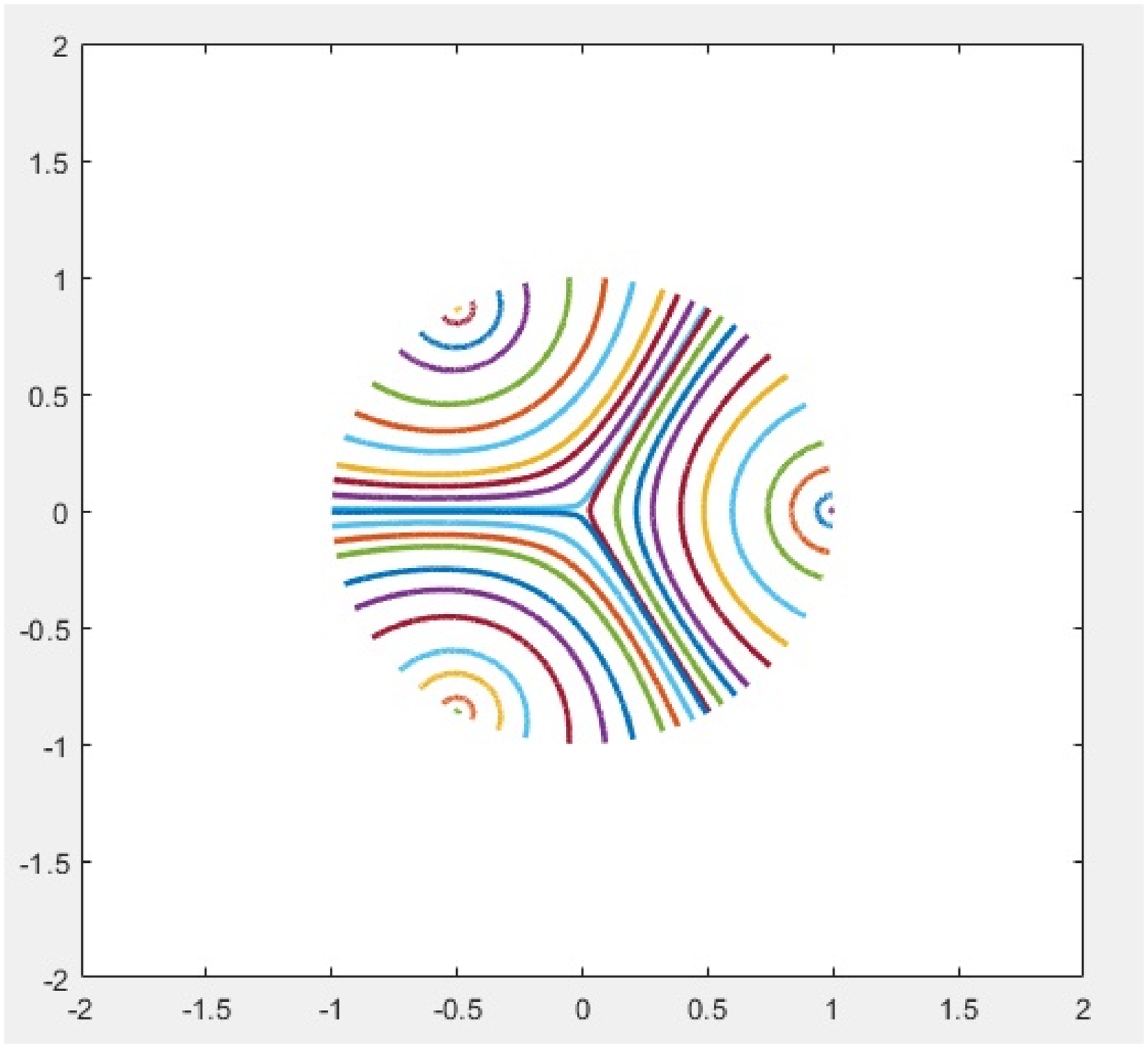}
\end {center}
\caption {\label{oplane} The three-open-string world-sheet mapped onto a unit disk on the $\o$-complex plane.}
\end{figure}

To evaluate the Polakov string path integral with temporal boundary on the string world-sheet, we need 
a Green's function for the coordinate fields $X^I$, $I = 0, \cdots,  d-1$. However, it is difficult to find a Green's 
function on the string world-sheet directly. Therefore, we map the string world-sheet to the upper half complex plane,
where Green's function is simple. 
First, we map the world-sheet onto a unit disk by a conformal transformation:
\beq\label{cs1}
\omega_1 &=& e^{\frac{2\pi i}{3}} \left(\frac{1+ i e^{\zeta_1}}{1 - i e^{\zeta_1}}\right)^{\frac{2}{3}}, \nn\\
\omega_2 &=& \left(\frac{1+ i e^{\zeta_2}}{1 - i e^{\zeta_2}}\right)^{\frac{2}{3}},\\
\omega_3 &=& e^{-\frac{2\pi i}{3}} \left(\frac{1+ i e^{\zeta_3}}{1 - i e^{\zeta_3}}\right)^{\frac{2}{3}}. \nn
\eeq 
where the local coordinates on the three patches are given as $\zeta_r = \xi_r + i \eta_r$, $r= 1, 2, 3$.
Fig. \ref{oplane} Depicts the three-open-string world-sheet mapped on a unit disk ($\o$-plane).   

\begin{figure}[htbp]
\begin {center}
\epsfxsize=0.5\hsize

\epsfbox{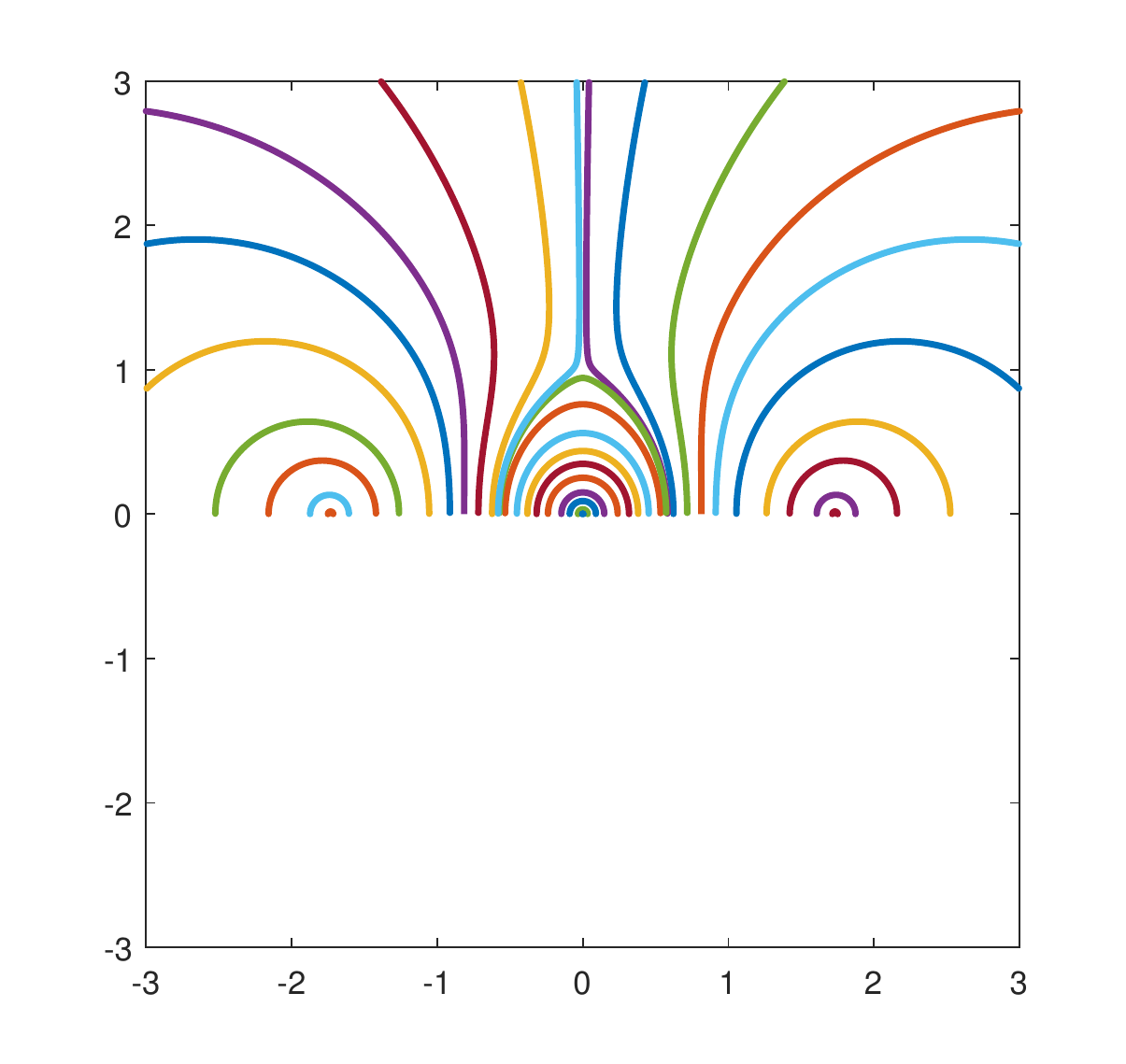}
\end {center}
\caption {\label{zplane} The world-sheet mapped onto the $z$-complex upper half-plane.}
\end{figure}

The interaction point $B$, where all three open strings meet, is mapped to the origin of the disk; the external strings are located at 
$e^{\frac{2\pi i}{3}}, ~ 1, ~ e^{-\frac{2\pi i}{3}}$. 
Then, each local coordinate patch on the unit disk is mapped onto the 
upper half-plane by the following conformal transformation:
\beq \label{cs2}
z = -i \,\frac{\omega_r -1}{\omega_r +1}, ~~~
\frac{\pi}{3} \le \arg\, \omega_r \le \frac{2\pi}{3}, ~~~ r =1, ~2,~ 3.
\eeq 
The three-open-string world-sheet mapped on the $z$-plane is described by Fig. \ref{zplane} \cite{Lee2020cubic}. 
The external strings are mapped to three points on the real line
\beq
Z_1 = \sqrt{3}, ~~ Z_2 = 0, ~~~ Z_3 = - \sqrt{3}.
\eeq 

Having mapped the world-sheet of three strings onto the upper half-plane, 
we can adopt the well-known Green's functions on the upper half complex plane,
\beq
G_N(z,\zp) &=& \ln \vert z- \zp \vert + \ln \vert z- z^{\prime *} \vert, ~~~\text{for} ~~\text{Neumann boundary condition},\nn\\
G_D(z,\zp) &=& \ln \vert z- \zp \vert - \ln \vert z- z^{\prime *} \vert, ~~~\text{for} ~~\text{Dirichlet boundary condition}.
\eeq 
In order to illustrate the procedure we shall follow, we consider the free string case first. The world-sheet of the free
open string is an infinite strip, which is mapped onto the upper half plane by a simple conformal transformation 
\beq
\zeta = \ln z,
\eeq  
where  $\zeta = \xi + i \eta$. 
The Green's function on the upper half complex plane in the Neumann direction (for $X^\m$) and the Green's function 
in the Dirichlet direction (for $X^i$), respectively, are given as follows:
\beq
N(\zeta, \zeta^\prime)
&=& \ln \vert e^\zeta - e^{\zeta^\prime}\vert - 
\ln \vert e^\zeta - e^{\zeta^{\prime *} }\vert \nn\\
&=& - \sum_{n \ge 1} \frac{2}{n} e^{-n\vert \xi_r - \xi^\prime_s \vert} \cos (n\eta_r) 
\cos (n \eta^\prime_s) - 2 \text{max} \left(\xi_r, \xi^\prime_s \right) \nn\\
 D(\zeta, \zeta^\prime)
 &=& \ln \vert e^\zeta - e^{\zeta^\prime}\vert - 
 \ln \vert e^\zeta - e^{\zeta^{\prime *} }\vert \nn\\
 &=& -\sum_{n=1} \frac{2}{n} e^{-n\vert \xi - \xi^\prime\vert} 
 \sin n \eta \sin n \eta^\prime .
 \eeq

In case of interacting strings, the Green's functions may be written as
 \begin{subequations}
\beq \label{neumanna1}
N(\r_r, \rp_s) &=& \ln \vert z - \zp \vert + \ln \vert z - \zp^* \vert \nn\\
&=& - \d_{rs} \Biggl\{ \sum_{n \ge 1} \frac{2}{n} e^{-n\vert \xi_r - \xi^\prime_s \vert} \cos (n\eta_r) 
\cos (n \eta^\prime_s) - 2 \text{max} \left(\xi_r, \xi^\prime_s \right) \Biggr\} \nn\\
&& + 2 \sum_{n, m \ge 0} \bar N^{rs}_{nm} e^{n \xi_r + m \xi^\prime_s} \cos (n\eta_r) \cos (m\eta^\prime_s), \nn\\
D(\rho_r, \rho^\prime_s) &=& \ln \vert z - \zp \vert - \ln \vert z - \zp^* \vert \\
&=& - \d_{rs} \Biggl\{\sum_{n\ge 1} \frac{2}{n} e^{-n\vert \xi_r - \xi^\prime_s \vert}
\sin\left(n\eta_r\right) \sin\left(n\eta^\prime_s\right)
\Biggr\} \nn\\
&&+ 2 \sum_{n, m \ge 0} \bar D^{rs}_{nm} e^{n\xi_r + m\xi^\prime_s} \sin\left(n\eta_r\right) \sin\left(m\eta^\prime_s\right) \label{dirichlet}
\eeq
\end{subequations}
where $\rho_r$ and $\rho^\prime_s$ lie in the region of the $r$-th and $s$-th local patches respectively.
The general formula for the Neumann functions $\bar N^{rs}_{nm}$ is given in the Appendix.  

Taking the limit $\zp \rightarrow Z_s$ ($\eta_s \rightarrow -\infty$) or $\zp \rightarrow Z_r$ ($\eta_r \rightarrow -\infty$) of Eq. (\ref{dirichlet}), 
we find 
\beq \label{Dn0}
\bar D^{rs}_{n0} = 0, ~~ \bar D^{rs}_{0n} = 0,~~~\text{for}~~ n \ge 0 . 
\eeq
By differentiating Eq. (\ref{dirichlet}) with respect to $\zeta_r$, we obtain 
\beq 
\bar D^{rs}_{nm} &=& - \frac{1}{nm} \oint_{Z_r} \frac{dz}{2\pi i} \oint_{Z_s} \frac{d z^\prime}{2\pi i} \frac{1}{(z-z^\prime)^2} e^{-n\zeta_r(z) - m \zeta^\prime_s(z^\prime)}, ~~~ n, m \ge 1. 
\eeq 
Hence, it turns out that 
\beq\label{Dnm}
\bar D^{rs}_{nm} = - \bar N^{rs}_{nm} , ~~\text{for}~~ n, m \ge 1.
\eeq 
It is interesting that we need only calculate the Neumann functions to construct the Fock space representations of the multi-string vertices on $Dp$-branes. 
Putting all this together, we may write the Fock space 
representation of the three-string vertex in terms of the Neumann function as
\beq 
E[1,2,\dots, N] \vert 0 \rangle
&=& \exp \,\Biggl\{ \sum_r \ln a_r \left(\frac{(p^{(r)})^2}{2}-1\right) \Biggr\} 
\prod_{r<s}\vert Z_r -Z_s \vert^{p^{(r)} \cdot p^{(s)}} \nn\\
&& \exp\Biggl\{ \frac{1}{2} \sum_{r,s =1}^3 \sum_{n, m \ge 1} \bar N^{rs}_{nm} \, 
\a^{(r)\dagger}_{n\m} \a^{(s)\dagger}_{m\n} \eta^{\m\n}  \nn\\
&& + \sum_{r=1}^3 \sum_{n \ge 1} \bar N^{rs}_{n0} \a^{(r)\dag}_{n\m} p^\m
- \frac{1}{2} \sum_{r,s =1}^3 \sum_{n, m \ge 1} \bar N^{rs}_{nm} \, 
\a^{(r)\dagger}_{n i} \a^{(s)\dagger}_{m j} \eta^{ij} \Biggr\} \vert 0 \rangle, \label{EN}.
\eeq
where the Schwarz-Christoffel (SC) map on the $r$-th local coordinate patch is defined as a series expansion
\beq 
e^{-\z_r} &=& \frac{a_r}{(z_r-Z_r)} + \sum_{n=0} c^{(r)}_n (z_r-Z_r)^n .
\eeq

\section{One-String Vertex for String Field Theory on Multiple $Dp$-branes}

The identity functional $I$ for the closed string with respect to $*$ may be given by an overlapping delta functional as 
\beq
I[X(\s)] = \langle X(\s) \vert I \rangle = \prod_{ 0 \le \s \le \pi/2} \d \left(X(\s) - X(\pi -\s) \right)
\eeq 
This defines the one-string vertex operator. A pictorial representation of 
the overlapping delta functional is presented in Fig..

\begin{figure}[htbp]
\begin {center}
\epsfxsize=0.6\hsize

\epsfbox{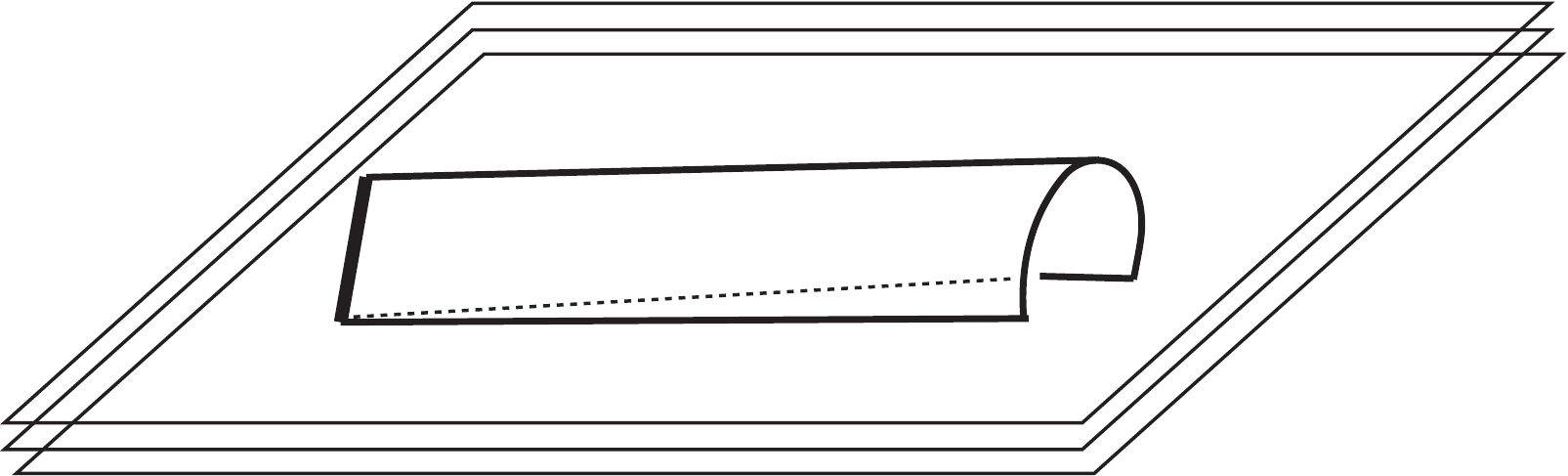}
\end {center}
\caption {\label{onestring} One-string vertex on $Dp$-branes.}
\end{figure}

We can choose the SC mapping from the string world-sheet to a unit disk  on the $\o$-plane as 
\beq
\omega &=& \left(\frac{1+ i e^{\zeta}}{1 - i e^{\zeta}}\right)^2, ~~~ 0 \le \eta \le \pi .
\eeq
On the $\omega$-plane, the external string is located at $\o=1$. 
The disk can be mapped onto the $z$-complex plane by a well-known conformal mapping
\beq
z = -i \frac{\o -1}{\o + 1}=-i \frac{\left(\frac{1+ i e^{\zeta}}{1 - i e^{\zeta}}\right)^2 -1}{\left(\frac{1+ i e^{\zeta}}{1 - i e^{\zeta}}\right)^2 + 1} .
\eeq
The external string is mapped onto $Z=0$ on the $z$-complex plane. 
It describes an open string, propagating with one end sealed by the overlapping condition. (See Fig. \ref{onestring}.) Thanks to momentum conservation, component fields do not carry a momentum. 

The Fock space representation of the one-closed-string identity (vertex) is obtained from expansion of $e^{-\z}$ around $Z=0$
\beq
e^{-\z} &=& - i \left(\frac{1+ \o^\half }{1-\o^\half} \right) \nn\\
&=& \frac{2}{z}+\frac{z}{2}-\frac{ z^3}{8}+\frac{z^5}{16}-\frac{5 z^7}{128}+\frac{7 z^9}{256}+O\left(z^{11}\right) \nn\\
&=& \frac{2}{z} + \sum_{n=0} c_n z^n .
\eeq 

The Fock representation of the one-string vertex follows from the details of calculation given in the Appendix 
\beq
\vert I_1 \rangle 
&=& \frac{1}{2} \exp  \Biggl\{
\frac{1}{2} \sum_{n, m \ge 1} \bar N_{nm}\,\a^{\dag}_n \cdot 
\a^{\dag}_m- \frac{1}{2} \sum_{n,m \ge 1} 
\bar N_{nm} \a^\dag_{ni}\a^\dag_{mj} \eta^{ij} 
\Biggr\}
\eeq  
where
\beq
\bar N_{00} &=& \ln 2 , ~~ \bar N_{10} = 0, ~~  \bar N_{20} =1, ~~ \bar N_{30} =0, ~~ \bar N_{40} = \half, ~~  \bar N_{50} =0,  \cdots, \nn\\
\bar N_{11} &=& 1 , ~~ \bar N_{12} = 0,~~\bar N_{21} = 0 , ~~ \bar N_{22} = 1,~~\bar N_{13} = 0 , ~~ \bar N_{23} = 0, \cdots . 
\eeq 

\section{Two-String Vertex for Cubic String Field Theory on Multiple $Dp$-branes}

Two-string overlapping, which defines the two-string vertex, may be written as 
\beq
\langle X^{(1)}, X^{(2)} \vert I \rangle &=& \prod_{\frac{\pi}{2} \le \s \le \pi}
\d \left(X^{(1)}(\s) - X^{(2)}(\pi - \s ) \right) \nn\\
&& 
\prod_{\frac{\pi}{2} \le \s \le \pi}
\d \left(X^{(2)}(\s) - X^{(1)}(\pi - \s ) \right) .
\eeq
The mapping from the world-sheet coordinates $\zeta_r = \xi_r + i \eta_r$, $r= 1, 2$ onto the unit
disk is given as follows: 
\begin{subequations}
\beq
\omega_1 &=&  -i \left(\frac{1+ i e^{\zeta_1}}{1 - i e^{\zeta_1}}\right) = -i\left(\frac{1 - e^{2{\xi_1}} + 2i e^{\xi_1} \cos\eta_1}{ 1 + e^{2\xi_1} + 2 e^{\xi_1} \sin^2\eta_1}\right) , \\
\omega_2 &=& i \left(\frac{1+ i e^{\zeta_2}}{1 - i e^{\zeta_2}}\right)= i \left(\frac{1 - e^{2\xi_2} + 2i e^{\xi_2} \cos\eta_2}{ 1 + e^{2\xi_2} + 2 e^{\xi_2}\sin^2\eta_2}\right).
\eeq 
\end{subequations}

\begin{figure}[htbp]
\begin {center}
\epsfxsize=0.6\hsize

\epsfbox{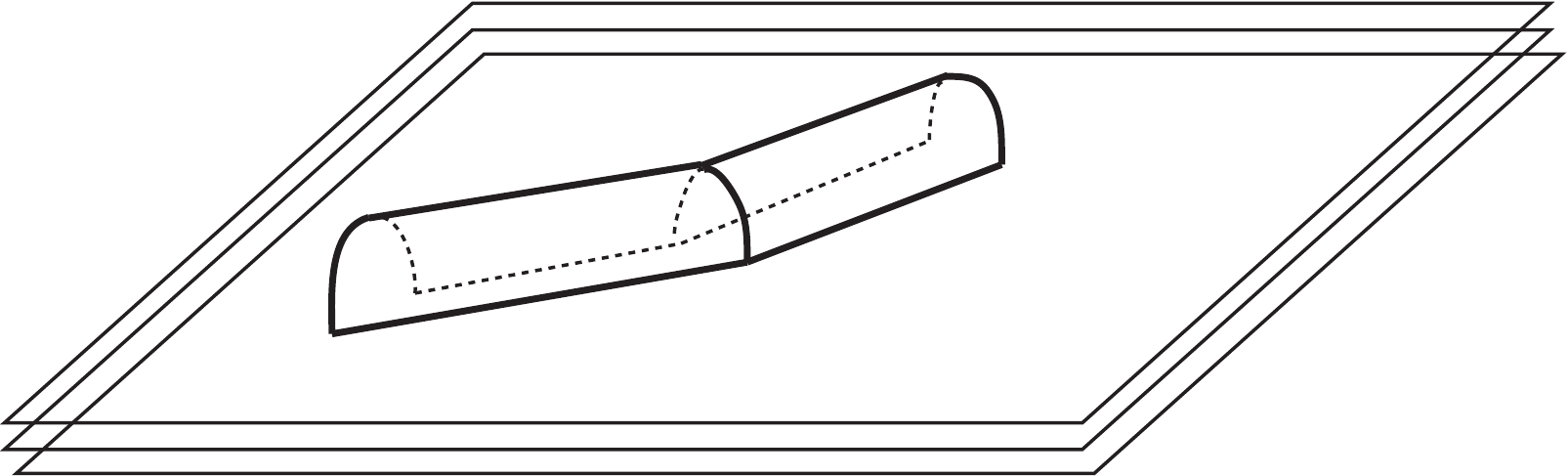}
\end {center}
\caption {\label{Dpbrane2} Two-string vertex on $Dp$-branes.}
\end{figure}

Two external strings (at asymptotic regions) are located at $-i$ and $i$ at the unit disk on the $\o$-complex plane respectively.
Since the ranges of local coordinates $\eta_r$, $r=1,2$ are confined to $[0,\pi]$, the images of string world trajectories are inside the unit disk: 
\beq
|\o_r| = \frac{ 1 + e^{2\xi_r} -2 e^{\xi_r} \sin \eta_r}{1 + e^{2\xi_r} +2 e^{\xi_r} \sin \eta_r},
\eeq 
This describes a linear coupling of two open strings on $Dp$-branes. 
This two-string vertex may be useful to study a linear coupling of open strings on $Dp$-branes 
of different kinds. 

Mapping onto the complex plane may be carried out as follows:
\beq
z_1 &=& -i \frac{\o_1 -1}{\o_1 + 1} =  -i \frac{ -i \left(\frac{1+ i e^{\zeta_1}}{1 - i e^{\zeta_1}}\right)  -1}{ -i \left(\frac{1+ i e^{\zeta_1}}{1 - i e^{\zeta_1}}\right)  + 1}, \nn\\
z_2 &=& -i \frac{\o_2 -1}{\o_2 + 1} = -i \frac{i \left(\frac{1+ i e^{\zeta_2}}{1 - i e^{\zeta_2}}\right) -1}{i \left(\frac{1+ i e^{\zeta_2}}{1 - i e^{\zeta_2}}\right) + 1} 
\eeq 
with $\o_r = \left(\frac{1 +iz_r}{1-iz_r} \right)$. 
This maps the external strings $Z_1 = -1 $ and $Z_2 =1$ on the real line of the $z$-complex plane
\beq \label{expand27}
e^{-\z_1} &=& - i \left(\frac{1+ i\o_1}{1-i \o_1} \right) =  - i \left(\frac{1+ i     \left(\frac{1 +iz_1}{1-iz_1} \right) }{1-i  \left(\frac{1 +iz_1}{1-iz_1} \right)} \right) = \frac{2}{z_1+1} -1, \nn\\
e^{-\z_2} &=& - i \left(\frac{1- i\o_2}{1+i \o_2} \right) =   - i \left(\frac{1 -i     \left(\frac{1 +iz_2}{1-iz_2} \right) }{1+ i  \left(\frac{1 +iz_2}{1-iz_2} \right)} \right) = \frac{2}{z_2-1}+1 . 
\eeq
As in the case of the one-string vertex operator, if the ranges of $\eta_r$ $r=1, 2$ are limited to $[0,\pi]$, the images of the 
string world-sheet cover only the upper half complex $z$-plane 

The Fock space representation of the two-closed-string (identity) vertex follows from the general expression of the vertex operator, Eq. (\ref{EN}) 
\beq
\vert I[2] \rangle &=&
\exp \Biggl\{ \sum_{r=1}^2 \ln 2 \left( \frac{\left(p^{(r)}\right)^2}{2} -1 \right) 
\Biggr\}~ 2^{ p^{(1)} \cdot p^{(2)}}  \nn\\
&&
\exp  \Biggl\{
\sum_{r,s} \Bigl( \sum_{n, m \ge 1} \frac{1}{2} \bar N^{rs}_{nm}\,\a^{(r)\dag}_n \cdot \a^{(r)\dag}_m 
+ \sum_{n \ge 1}\bar N^{rs}_{n0} \a^{(r)\dag}_n \cdot p^{(s)} \Bigr) \Biggr\} \nn\\
&& \exp  \Biggl\{
-\sum_{r,s} \Bigl( \sum_{n, m \ge 1} \frac{1}{2} \bar N^{rs}_{nm}\,\a^{(r)\dag}_{ni} \a^{(r)\dag}_{mj}\eta^{ij} \Bigr) \Biggr\}
\vert 0 \rangle . 
\eeq 
Here the Neumann functions are explicitly evaluated as 
\beq
\bar N^{12}_{00} &=&   \ln \vert Z_1 - Z_2 \vert = \ln 2, ~~\bar N^{11}_{00} = \bar N^{22}_{00} = \ln 2\nn\\
N^{11}_{10} &=& -1, ~~N^{22}_{10} =1, ~~ N^{12}_{10}= -1, ~~ N^{21}_{10} = 1,\nn\\
N^{11}_{20} &=& \half, ~~N^{22}_{20} =\half, ~~ N^{12}_{20}= 1, ~~ N^{21}_{20} = 1,\nn\\
N^{11}_{30} &=& - \frac{1}{3}, ~~N^{22}_{30} =\frac{1}{3}, ~~ N^{12}_{30}= -1, ~~ N^{21}_{30} = 1.
\eeq 

Using Eq. (\ref{expand27}) and the general formula for the Neumann function $\bar N^{rs}_{nm}$, $n,m \ge 1$ given by Eq. (\ref{Nnm}), we are able to calculate $\bar N^{rs}_{11}$ for the Neumann function of the two-string vertex, which we will need shortly:
\begin{subequations}
\beq
\bar N^{11}_{11} &=& \oint_{Z_1} \frac{dz_1}{2\pi} \oint_{Z_1} \frac{dz^\prime_1}{2\pi} \frac{1}{(z_1-z^\prime_1)^2} \left(\frac{2}{z_1+1} -1\right)\left(\frac{2}{z^\prime_1+1} -1\right) = 0, \\
\bar N^{12}_{11} &=& \oint_{Z_1} \frac{dz_1}{2\pi} \oint_{Z_2} \frac{dz^\prime_2}{2\pi} \frac{1}{(z_1-z^\prime_2)^2} \left(\frac{2}{z_1+1} -1\right)\left(\frac{2}{z^\prime_2-1} -1\right) = \frac{1}{2}, \\
\bar N^{21}_{11} &=& \oint_{Z_2} \frac{dz_2}{2\pi} \oint_{Z_1} \frac{dz^\prime_1}{2\pi} \frac{1}{(z_2-z^\prime_1)^2} \left(\frac{2}{z_2-1} -1\right)\left(\frac{2}{z^\prime_1+1} -1\right) = \frac{1}{2},\\
\bar N^{22}_{11} &=& \oint_{Z_2} \frac{dz_2}{2\pi} \oint_{Z_2} \frac{dz^\prime_2}{2\pi} \frac{1}{(z_2-z^\prime_2)^2} \left(\frac{2}{z_2-1} -1\right)\left(\frac{2}{z^\prime_2-1} -1\right) = 0.
\eeq 
\end{subequations}
We can make use of the two-string vertex to evaluate the quadratic term in the string field action in terms of component fields. Let us choose 
the external string state to calculate the string field action in the low energy region as follows: 
\beq
\langle \Psi^{(1)}, \Psi^{(2)} \vert&=& \Bigl\langle 0 \Bigl\vert \prod_{r=1}^2\Bigl(A_\m(p^{(r)}) a^{(r)}_{1\n} \eta^{\m\n}
+ \varphi_i(p^{(r)}) a^{(r)}_{1j} \eta^{ij} \Bigr ) \nn\\
&=& \Bigl\langle 0 \Bigl\vert \prod_{r=1}^2\Bigl( \bolA(r) + \bolvarphi(r) \Bigr)^2.
\eeq 
The quadratic terms in the string field action may be written in the low energy region as 
\beq
{S}_{[2]}
&=&  g_2\int \prod_{r=1}^2 dp^{(r)} \d \left(\sum_{r=1}^2 p^{(r)} \right) \,\text{tr}\,\Bigl\langle 0 \Bigl\vert \prod_{r=1}^3\Bigl( \bolA(r) + \bolvarphi(r) \Bigr)^2 \Bigr\vert I_2 \Bigr\rangle \nn\\
&=& g_2\int \prod_{r=1}^2 dp^{(r)} \d \left(\sum_{r=1}^2 p^{(r)} \right) \,\text{tr}\,\Bigl\langle 0 \Bigl\vert \Bigl(\bolA(1) \bolA(2) +
\bolA(1)\bolvarphi(2) \nn\\
&& + \bolvarphi(1) \bolA(2) + \bolvarphi(1) \bolvarphi(2) \Bigr) \Bigr\vert I_2 \Bigr \rangle .
\eeq
The quadratic terms in the gauge field may be read as 
\beq
&& g_2\int \prod_{r=1}^2 dp^{(r)} \d \left(\sum_{r=1}^2 p^{(r)} \right) \,\text{tr}\,\bigl\langle 0 \bigl\vert \bolA(1)\bolA(2) + \bolA(2)\bolA(1) \bigr\vert I_2 \bigr \rangle \nn\\
&=& g_2 \int \prod_{r=1}^2 dp^{(r)} \d \left(\sum_{r=1}^2 p^{(r)} \right) \bigl\langle 0 \bigl\vert \,\text{tr}\left\{ A_\m(1) a^{(1)\m}_1 A_\n(2) a^{(2)\n}_1+ A_\n(2) a^{(2)\n}_1 A_\m(1) a^{(2)\m}_1  \right\} \nn\\
&& \exp\left\{ \sum_{r,s}  \frac{1}{2} \bar N^{rs}_{11}\,\a^{(r)\dag}_1 \cdot \a^{(r)\dag}_1   \right\}\bigr \vert  0 \bigr \rangle \nn\\
&=&  g_2\int \prod_{r=1}^2 dp^{(r)} \d \left(\sum_{r=1}^2 p^{(r)} \right) \left\{\frac{1}{2} A(1)\cdot A(2) \bar N^{12}_{11} + \frac{1}{2} A(2)\cdot A(1) \bar N^{21}_{11} \right\} \nn\\
&=& \frac{g_2}{2} \int d^{p+1} x \, {\rm tr} \, A^2 . 
\eeq
As we expect, in the low energy region, the quartic string field term yields a
mass term for the gauge field. The gauge-scalar-field linear coupling simply vanishes: 
\beq
\text{tr}\,\Bigl\langle 0 \Bigl\vert \Bigl(\bolA(1)\bolvarphi(2)  + \bolvarphi(1) \bolA(2)  \Bigr) \Bigr\vert I_2 \Bigr \rangle =0.
\eeq
We anticipate that the quadratic term of the scalar field also reduces to a mass term for the scalar field: 
\beq
 &&g_2\int \prod_{r=1}^2 dp^{(r)} \d \left(\sum_{r=1}^2 p^{(r)} \right) \,\text{tr}\,\Bigl\langle 0 \Bigl\vert\bolvarphi(1) \bolvarphi(2) \Bigr) \Bigr\vert I_2 \Bigr \rangle \nn\\
 &=& g_2\int \prod_{r=1}^2 dp^{(r)} \d \left(\sum_{r=1}^2 p^{(r)} \right)
 \,\text{tr}\,\Bigl\langle 0 \Bigl\vert \varphi_i(1) a^{(1)}_{1j} \eta^{ij}
 \varphi_k(2) a^{(2)}_{1l} \eta^{kl} \nn\\
 &&  \exp\left\{ - \sum_{r,s} \frac{1}{2} \bar N^{rs}_{11} a^{(r)\dag}_{1p} a^{(s)\dag}_{1q} \eta^{pq} \right\} \Bigr\vert 0 \Bigr \rangle \nn\\
 &=& - \frac{g_2}{2} \int \prod_{r=1}^2 dp^{(r)} \d \left(\sum_{r=1}^2 p^{(r)} \right) \,\text{tr}\,\Bigl\langle 0 \Bigl\vert \varphi_1(1) \varphi_i (2) \bar N^{12}_{11} + \varphi_i(2) \varphi_i(1) \bar N^{21}_{11}  \Bigr\vert 0 \Bigr \rangle \nn\\
 &=& - \frac{g_2}{2} \int d^{p+1} x~ {\rm tr}~\varphi^2 . 
\eeq 
However, the signature of the quadratic term of the scalar is negative (if the 
coupling constants $g_2$ is positive). This implies that the quartic string field term on multiple $Dp$-branes gives a negative mass to the scalar fields in the low energy region.

\section{Three-String Vertex for Cubic String Field Theory on Multiple $Dp$-branes}

\begin{figure}[htbp]
\begin {center}
\epsfxsize=0.6\hsize

\epsfbox{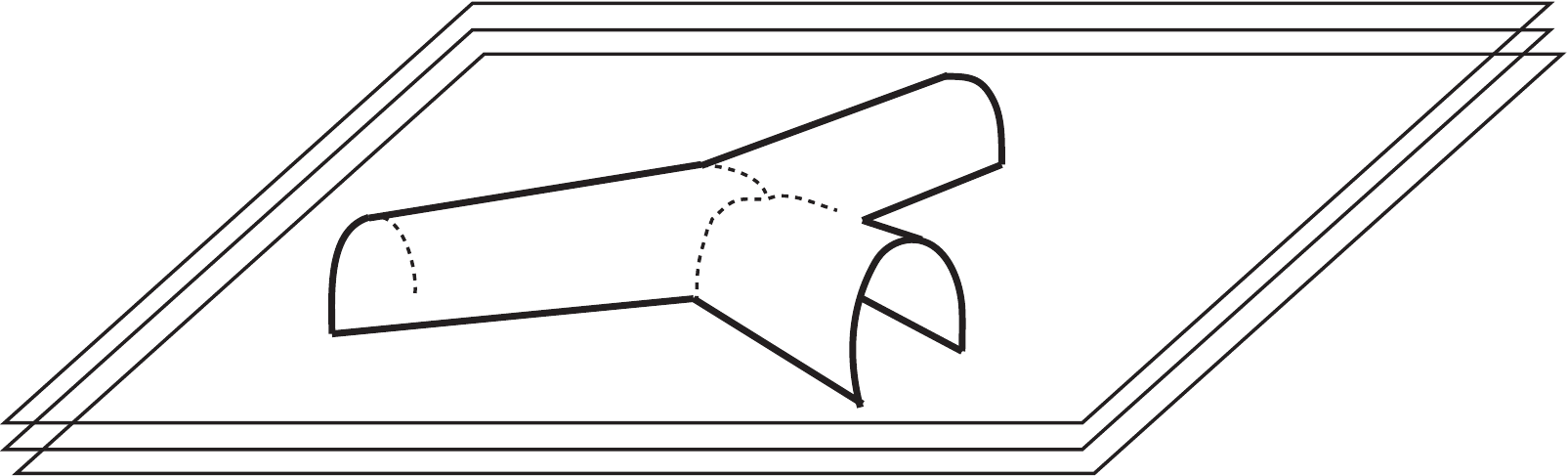}
\end {center}
\caption {\label{Dpbrane3} Three-string vertex on $Dp$-branes.}
\end{figure}

The cubic open string with the 
star product Eq. (\ref{star}) defined by Witten is pictorially depicted by Fig. \ref{Dpbrane3}. 
The mapping from the world-sheet coordinates $\zeta_r = \xi_r + i \eta_r$, $r= 1, 2, 3$ to the 
disk is given by Eq. (\ref{cs1}). We employ the following conformal transformation to map the 
complex $\o$-plane to the complex $z$-plane: 
\beq
z_r = -i \,\frac{\omega_r -1}{\omega_r +1}, ~~ r =1, ~2,~ 3.
\eeq 
The external strings are now located on the real line 
\beq
Z_1 = \sqrt{3}, ~~ Z_2 = 0, ~~~ Z_3 = - \sqrt{3}.
\eeq 
Each local coordinate patch is mapped onto the upper half plane for 
$ 0 \le \eta_r \le \pi, ~~r = 1, 2, 3$ (See Fig. \ref{zplane}.)	

The relations between the local coordinates and $z$-complex coordinates may be manifested through expansions of  $e^{-\z_r}$, $r=1,2,3$ near $z_r = Z_r$ (at the asymptotic region),
\beq
e^{-\z_r} &=& \frac{a_r}{(z_r-Z_r)} + \sum_{n=0} c^{(r)}_n (z_r-Z_r)^n , \label{expand46}\\
a_1 &=& \frac{8}{3}, ~~ a_2 = \frac{2}{3}, ~~~ a_3 = \frac{8}{3}, \nn\\
c^{(1)}_0 &=&\frac{2\sqrt{3}}{3}, ~~ c^{(1)}_1 = - \frac{5}{72} , ~~c^{(1)}_2 = \frac{5\sqrt{3}}{288} \nn\\
c^{(2)}_0 &=& 0, ~~ c^{(2)}_1 = - \frac{5}{18}, ~~ c^{(2)}_2 = 0 ,\nn\\
c^{(3)}_0 &=& -\frac{2\sqrt{3}}{3}, ~~ c^{(3)}_1 = - \frac{5}{72} , ~~c^{(3)}_2 = -\frac{5\sqrt{3}}{288} . \nn 
\eeq
These explicit expressions of expansions are useful for calculating the Neumann functions.

to Calculating them explicitly involves some algebra. We evaluate Neumann functions of $\bar N^{rs}_{00}$ using Eq. (\ref{expand46}) and Eq. (\ref{Nrr00}) given in the Appendix:
\beq \label{N00}
\bar N^{11}_{00} &=& \ln \frac{8}{3}, ~~\bar N^{22}_{00} = \ln \frac{2}{3}, ~~ \bar N^{33}_{00} = \ln \frac{8}{3} ,\nn\\
\bar N^{12}_{00} &=& \frac{1}{2}\ln 3, ~~\bar N^{23}_{00} = \frac{1}{2} \ln 3, ~~ \bar N^{31}_{00} =  \ln 2 \sqrt{3} . 
\eeq 
The Neumann functions of $\bar N^{rs}_{10}$ follow from Eq. (\ref{expand46}) and Eq. (\ref{Nn0}):
\beq
N^{rr}_{10} = c^{(r)}_0   ,~~~\bar N^{rs}_{10} = \frac{a_r}{(Z_r-Z_s)}, ~~\text{for}~~r\not=s . 
\eeq 
Spelling these out, we obtain: 
\beq \label{N10}
\bar N^{11}_{10} &=&\frac{2\sqrt{3}}{3}, ~~\bar N^{22}_{10}= 0 , ~~\bar N^{33}_{10}= - \frac{2\sqrt{3}}{3}, \nn\\ 
\bar N^{12}_{10} &=& \frac{8}{3\sqrt{3}}, ~~ \bar N^{13}_{10} =  \frac{4}{3\sqrt{3}}, ~~
\bar N^{21}_{10} = - \frac{2}{3 \sqrt{3}}, \\
\bar N^{23}_{10} &=& \frac{2}{3\sqrt{3}}, ~~ \bar N^{31}_{10} = -  \frac{4}{3\sqrt{3}}, ~~
\bar N^{32}_{10} = - \frac{8}{3\sqrt{3}} .  \nn
\eeq

The Neumann functions of $\bar N^{rs}_{11}$ may be obtained from the general formula Eq. (\ref{Nnm}) for $\bar N^{rs}_{nm}$, given in the Appendix:
\beq \label{Nnm}
\bar N^{rs}_{mn} =\frac{1}{nm} \oint_{Z_r} \frac{dz_r}{2\pi i} \oint_{Z_s} \frac{d z^\prime_s}{2\pi i} 
\frac{1}{(z_r-z^\prime_s)^2} e^{-n\zeta_r(z_r) - m \zeta^\prime_s(z^\prime_s)}, ~~~ n, m \ge 1
\eeq 
For the case with $n=1$ and $m=1$:
\beq
\bar N^{rs}_{11} &=& \oint_{Z_r} \frac{dz_r}{2\pi i} \oint_{Z_s} \frac{d z^\prime_s}{2\pi i} 
\frac{1}{(z_r-z^\prime_s)^2} \left( \frac{a_r}{z_r-Z_r}\right)\left( \frac{a_s}{z^\prime_s-Z_s}\right). 
\eeq 
When $r\not= s$, 
\beq
\bar N^{rs}_{11} &=& \frac{a_r a_s}{(Z_r -Z_s)^2} = \frac{2^4}{3^3} , ~~~{\rm for}~~ r, s = 1,2,3.
\eeq 
When $r=s$, we may write 
\beq
\bar N^{rr}_{11} &=& \oint_{Z_r} \frac{dz_r}{2\pi i} \oint_{Z_s} \frac{d z^\prime_r}{2\pi i} 
\frac{1}{(z_r-z^\prime_r)^2} \left( \frac{a_r}{z-Z_r}+ \sum_{m=0} c^{(r)}_m (z_r-Z_r)^m \right) \nn\\
&&
\left( \frac{a_r}{z^\prime_r-Z_r}+ \sum_{m=0} c^{(r)}_m (z_r-Z_r)^m \right) \nn\\
&=& a_r c^{(r)}_1.
\eeq 
Using Eq. (\ref{expand46}) and 
\beq
c^{(1)}_1 &=& - \frac{5}{72}, ~~c^{(2)}_1 = - \frac{5}{18},~~ c^{(3)}_1 = - \frac{5}{72}, 
\eeq 
we find 
\beq
\bar N^{11}_{11} =\bar N^{22}_{11} =\bar N^{33}_{11} = - \frac{5}{27}. 
\eeq

With the Neumann functions for three closed strings Eq. (\ref{N00}), Eq. (\ref{N10}), and Eq. (\ref{Nnm}), the three-closed-string vertex operator may be expressed as 
\beq \label{vertexopen3}
E[1,2,3] \vert 0 \rangle 
&=&  \exp \Biggl\{ \sum_{r=1}^3 \ln \frac{8}{3} \left( \frac{\left(p^{(r)}\right)^2}{2} -1 \right) 
\Biggr\}\prod_{r <s} \vert Z_r - Z_s \vert^{ p^{(r)} \cdot p^{(s)}}~\nn\\
&&
\exp  \Biggl\{
\sum_{r,s} \Bigl( \sum_{n, m \ge 1} \frac{1}{2} \bar N^{rs}_{nm}\,\a^{(r)\dag}_n 
\a^{(s)\dag}_m
+ \sum_{n \ge 1}\bar N^{rs}_{n0} \a^{(r)\dag}_n \cdot p^{(s)} \Bigr) 
\Biggr\} \nn\\
&& \exp  \Biggl\{-\sum_{r,s} \Bigl( \sum_{n, m \ge 1} \frac{1}{2} \bar N^{rs}_{nm}\,\a^{(r)\dag}_{ni}\a^{(s)\dag}_{mj} \eta^{ij}\Bigr) \Biggr\} \vert 0 \rangle
\eeq


\section{Cubic String Field Theory in the Zero-slope Limit and Matrix Models}

The three-string interaction may be written as 
\beq \label{3string}
{S}_{[3]} = \frac{2g}{3} \int \prod_{r=1}^3 dp^{(r)} \d \left(\sum_{r=1}^3 p^{(r)} \right)  \langle \Psi_1, \Psi_2, \Psi_3 \vert E[1,2,3] \vert 0 \rangle.
\eeq 
In the low energy regime (or in the zero-slope limit), the external string states correspond to massless gauge fields $A^\m$ or massless scalar fields $\varphi^i$. By choosing the external string state as follows: 
\beq
\langle \Psi^{(1)}, \Psi^{(2)}, \Psi^{(3)} \vert&=& \Bigl\langle 0 \Bigl\vert \prod_{r=1}^3\Bigl(A_\m(p^{(r)}) a^{(r)}_{1\n} \eta^{\m\n}
+ \varphi_i(p^{(r)}) a^{(r)}_{1j} \eta^{ij} \Bigr ) \nn\\
&=& \Bigl\langle 0 \Bigl\vert \prod_{r=1}^3\Bigl( \bolA(r) + \bolvarphi(r) \Bigr)^3 ,
\eeq

We can evaluate the effective interaction between the gauge fields $A^\m$ and the scalar fields $\varphi^i$,
which describes the three-string interaction Eq. (\ref{vertexopen3}) and Eq. (\ref{3string}) in the zero-slope limit:
\beq \label{s3low}
{S}_{[3]}
&=&  \frac{2g}{3}\int \prod_{r=1}^3 dp^{(r)} \d \left(\sum_{r=1}^3 p^{(r)} \right) \,\text{tr}\,\Bigl\langle 0 \Bigl\vert \prod_{r=1}^3\Bigl( \bolA(r) + \bolvarphi(r) \Bigr)^3 E[1,2,3] \Bigr\vert 0 \Bigr\rangle \nn\\
&=& \frac{2g}{3}\int \prod_{r=1}^3 dp^{(r)} \d \left(\sum_{r=1}^3 p^{(r)} \right) \,\text{tr}\,\Bigl\langle 0 \Bigl\vert \Bigl(\bolA(1) \bolA(2) \bolA(3) +
\bolA(1)\bolvarphi(2)\bolvarphi(3) + \nn\\
&& + \bolvarphi(1) \bolA(2) \bolvarphi(3) + \bolvarphi(1) \bolvarphi(2) \bolA(3) \Bigr) E[1,2,3] \Bigr\vert 0 \Bigr \rangle
\eeq
Note that the $\varphi^3$ term vanishes because the vertex operator generates
only even powers of $a^{(r)}_{ni}$ : 
\beq
{\rm tr} \bigl \langle 0 \bigl\vert \bolvarphi(1) \bolvarphi(2) \bolvarphi(3) \bigr\vert 0 \bigr \rangle = 0 . 
\eeq

\subsection{Cubic Gauge Field Interaction}

The cubic gauge field interaction term can be written as 
\beq
S_{AAA}&=&\frac{2g}{3}\int \prod_{r=1}^3 dp^{(r)} \d \left(\sum_{r=1}^3 p^{(r)} \right) \,\text{tr}\,\Bigl\langle 0 \Bigl\vert \Bigl(\bolA(1) \bolA(2) \bolA(3)
E[1,2,3] \Bigr\vert 0 \Bigr \rangle \nn\\
&=&  \frac{2g}{3
}  \int \prod_{i=1}^3 d p^{(i)} \d \left(\sum_{i=1}^3 p^{(i)} \right)
\text{tr} \Bigl\langle 0 \Bigl \vert \left\{\prod_{i=1}^3 A(p^{(i)}) \cdot a^{(i)}_1 \right\} \nn\\
&& \left\{\frac{1}{2}\sum_{r, s =1}^3\bar N^{rs}_{11} \left(a^{(r)\dag}_1 \cdot a^{(s)\dag}_1 \right)\right\} \left\{\sum_{l,m=1}^3 \bar N^{lm}_{10} \left(a^{(l)\dag}_1 \cdot p^{(m)} \right)\right\}\Bigl \vert 0 \Bigl \rangle \nn\\
&=& \left(\frac{2}{3}\right)^4 g \int \prod_{i=1}^3 d p^{(i)} \d \left(\sum_{i=1}^3 p^{(i)} \right)\text{tr} \left(A(1)_\m A(2)_\n A(3)_\eta\right) \nn\\
&& \Biggl\{\eta^{\m\n} \frac{4}{\sqrt{3}} \left(p^{(3)}_\eta-p^{(2)}_\eta \right) +\eta^{\m\eta} \frac{2\sqrt{3}}{3}  \left(p^{(3)}_\n - p^{(1)}_\n\right)  +\eta^{\n\eta} \frac{4\sqrt{3}}{3} \left(p^{(2)}_\m-p^{(1)}_\m\right)
\Biggr\} \nn\\
&=& \frac{2^5}{3^5\sqrt{3}}~g \int \prod_{i=1}^3 d p^{(i)} \d \left(\sum_{i=1}^3 p^{(i)} \right)\,\eta^{\m\n} \left(p^{(1)}_\eta - p^{(2)}_\eta\right) {\rm tr}  \left(A(1)_\m A(2)_\n A(3)_\eta\right)               \nn\\
&=& g_{YM} \int \prod_{i=1}^3 d p^{(i)} \d \left(\sum_{i=1}^3 p^{(i)} \right)\, p^{(1)\m} \text{tr} \left(A(1)^\n [ A(2)_\n, A_\m(3)] \right)\nn\\
&=& - g_{YM} \int d^{p+1}x \, i\, {\rm tr}\, \left(\p_\m A_\n - \p_\n A_\m \right)[A^\m, A^\n ],
\eeq 
where 
$g_{YM} = \frac{2^5}{3^5\sqrt{3}}~g $. 
This term is precisely the cubic term in the covariant non-Abelian gauge action ${\rm tr} \,F_{\m\n} F^{\m\n}$

\subsection{Gauge-Scalar Field Interaction}

We may write the gauge-scalar cubic interaction term as follows:
\beq
S_{A\varphi\varphi} &=& \frac{2g}{3}\int \prod_{r=1}^3 dp^{(r)} \d \left(\sum_{r=1}^3 p^{(r)} \right) \,\text{tr}\,\Bigl\langle 0 \Bigl\vert \Bigl(
\bolA(1)\bolvarphi(2)\bolvarphi(3) + \nn\\
&& + \bolvarphi(1) \bolA(2) \bolvarphi(3) + \bolvarphi(1) \bolvarphi(2) \bolA(3) \Bigr) E[1,2,3] \Bigr\vert 0 \Bigr \rangle \nn\\
&=& \frac{2g}{3}\int \prod_{r=1}^3 dp^{(r)} \d \left(\sum_{r=1}^3 p^{(r)} \right) \,\text{tr}\,\Bigl\langle 0 \Bigl\vert \Bigl\{ A(1)_\m a^{(1)\m}_1 \varphi(2)_i a^{(2)i}  \varphi(3)_j a^{(3)j}  \nn\\
&&+ \varphi(1)_i a^{(1)i} A(2)_\m a^{(2)\m}_1 \varphi(3)_j a^{(3)j}+\varphi(1)_i a^{(1)i}  \varphi(2)_j a^{(2)j}A(3)_\m a^{(3)\m}_1  \Bigr\} \nn\\
&& \left\{\sum_{r,s=1}^3 \bar N^{rs}_{10} a^{(r)\dag}_n \cdot p^{(s)} \right\}
 \Biggl\{-\sum_{r,s} \Bigl( \frac{1}{2} \bar N^{rs}_{11}\, a^{(r)\dag}_{1i}   a^{(s)\dag}_{1j} \eta^{ij}\Bigr) \Biggr\} \Bigr\vert 0 \Bigr \rangle \nn\\
 &=& - \frac{2g}{3} \left(\frac{1}{2}\right) \frac{2^4}{3^3} \int \prod_{r=1}^3 dp^{(r)} \d \left(\sum_{r=1}^3 p^{(r)} \right) \,\left\{\bar \varphi(2)_i \varphi(3)_i A(1) \cdot \sum_s N^{1s}_{10} p^{(s)}  \right.\nn\\
 && \left.
 + \varphi(3)_i \varphi(1)_i A(2) \cdot \sum_s N^{2s}_{10} p^{(s)} +\varphi(1)_i \varphi(2)_i A(3) \cdot \sum_s N^{3s}_{10} p^{(s)}  \right\} \nn\\
&=&  - \frac{2g}{3} \left(\frac{1}{2}\right) \frac{2^4}{3^3} \int \prod_{r=1}^3 dp^{(r)} \d \left(\sum_{r=1}^3 p^{(r)} \right) \,\left\{\bar \varphi_i(2)\varphi(3)_i A(1) \cdot \frac{4}{3\sqrt{3}}(p^{(2)}-p^{(1)}) \right. \nn\\
&&\left .+ \varphi_i(3) \varphi(1)_i A(2) \cdot\frac{2}{3\sqrt{3}}(p^{(3)}-p^{(1)}) +\varphi(1)_i \varphi(2)_i A(3) \cdot \frac{4}{3\sqrt{3}}(p^{(3)}-p^{(2)})  \right\} \nn\\
&=& g_{YM}\,\int \prod_{r=1}^3 dp^{(r)} \d \left(\sum_{r=1}^3 p^{(r)} \right) \,p^{(1)\m} \, {\rm tr} \left( \varphi_i{(1)} [A{(3)}_\m, \varphi{(2)}_i ] \right) \nn\\
&=& -g_{YM} \int d^{p+1}x\, i\, {\rm tr} \,\p_\m \varphi_i [A_\m, \varphi_i].
\eeq
We confirm that this is the covariant gauge-scalar cubic coupling.

\section{Discussions and Conclusions}

We study Witten's string cubic open string field theory on multiple $Dp$-branes \cite{Dai1989,Horava1989,Polchinski1995}. On multiple $Dp$-branes the string field carries $U(N)$ group indices and its low energy excitations are massless non-Abelian gauge fields and massless scalar fields carrying $U(N)$ group indices. 
$Dp$-branes play important roles in string theory: They are dynamical objects with both ends of open strings attached, giving them non-Abelian symmetry. They appear as 
$p$-brane solutions in supergravity and provide a theoretical background for the holographic principle, which now has a wide range of applications in various fields. Thus, it is important to develop string field theory on multiple $Dp$-branes. Recently we studied covariant string field theory on multiple $Dp$-branes \cite{TLee2017cov} in the proper-time gauge \cite{Lee1988Ann}, which may be deformable to Witten's cubic string field theory \cite{Lai2018,Lee2019PLBfour}, and obtained some fruitful results: On multiple $D0$-branes string field theory reduces the
Bank-Fishler-Shenker-Susskind (BFSS) matrix model \cite{BFSS} multiple $D(-1)$-branes (D-instantons) to the Ishibashi-Kawai-Tsuchiya (IKKT) matrix model \cite{Ishibashi1997}. Since covariant string field theory in the proper-time gauge is deformable to Witten's cubic string field theory, we expect that Witten's cubic string field theory produces the same results. 

In the case of Witten's cubic string field theory, the SC mapping from the world-sheet to the complex $z$-plane is constructed in two steps: First we map the 
string world-sheet to a unit disk. Secondly, using a conformal transformation we map the unit disk to the upper half complex plane. Then the SC mapping from the 
string world sheet to the upper half complex plane can be expanded as a power series. We applied this procedure to one, two and three-string interaction to obtain vertex operators. The resultant Fock space representation of vertex operators is not exactly the same as the Fock space representation of overlapping conditions in the configuration space. The reason is that a string is not a point object: Right after interaction (overlapping) the string does not propagate completely freely. The temporal boundaries fixed by overlapping conditions still affect propagation of a string even when it  leaves the overlapping region. The correct vertex operator must 
be obtained from evaluating corresponding Polakov path integral. 

We have succeeded in constructing Fock space representation of one, two and three-string vertex operators for Witten's cubic open string field theory on multiple $Dp$-branes. However, we do not yet have a concrete proposal for a method to construct a
vertex operator of more than three strings for Witten's cubic string field theory. This is primarily due to our lack of understanding of conformal SC mapping from the world-sheet of four strings, which has two conical singular points, to the upper half complex plane. One plausible proposal would be to adopt deformation of the string world-sheet such as the four-string scattering diagram in the proper-time gauge
\cite{TLee2017cov}: If we deform the four-string scattering diagram to that in the proper-time gauge, the string world-sheet becomes planar, whereby the Mandelstam technique to construct a vertex operator is readily applicable. 

This work can be extended in diverse directions: We may define cubic string field theory on more complex D-brane systems such as parallel D-branes (separated) \cite{Karcz2012}, D1-D5 systems \cite{Seiberg1999}, and Non-BPS D-branes \cite{Sen2000cl}. Therefore, string field theory 
can be used as a useful tool for exploring important subjects associated with 
these D-brane systems, including tachyon condensation.

\vskip 1cm

\begin{acknowledgments}

This work was supported by a National Research Foundation of Korea (NRF) grant
funded by the Korean government (MSIT) (2021R1F1A106299311). 
\end{acknowledgments}

\begin{appendix}

\section{Calculations of the Neumann Functions of Open Strings on multiple $Dp$-branes }

The Fourier components of the Neumann function $N(\rho, \rho^\prime)$ are
\beq \label{neumanna1}
N(\r_r, \rp_s) &=& \ln \vert z - \zp \vert + \ln \vert z - \zp^* \vert \nn\\
&=& - \d_{rs} \left\{ \sum_{n \ge 1} \frac{2}{n} e^{-n\vert \xi_r - \xi^\prime_s \vert} \cos (n\eta_r) 
\cos (n \eta^\prime_s) - 2 \text{max} \left(\xi_r, \xi^\prime_s \right) \right\} \nn\\
&& + 2 \sum_{n, m \ge 0} \bar N^{rs}_{nm} e^{n \xi_r + m \xi^\prime_s} \cos (n\eta_r) \cos (m\eta^\prime_s)
\eeq
where $\rho_r$ and $\rho^\prime_s$ lie in the region of the $r$-th and $s$-th local patches respectively, and $\z_r = \xi_r + i \eta_r$, $r=1,2,3$ represents the local coordinates on the $r$-th local patch of the string world-sheet. 

The Fourier expansion of the Neumann function may also be written as 
\beq
N(\r_r, \rp_s) &=& - \d_{rs} \left\{\sum_{n \ge 1} \frac{1}{2n} \left(\o^{-n}_+ + \o^{*-n}_+\right)
\left(\o^{n}_- + \o^{*n}_-\right) - 2\, \text{max} (\xi_r, \xi^\prime_s)\right\} \nn\\
&& + \frac{1}{2} \sum_{n, m \ge 0} \bar N^{rs}_{nm} \left(\o^n_r + \o^{*n}_r\right) 
\left(\o^{\prime m}_s + \o^{\prime * m}_s\right) , \label{Nexpan2}
\eeq 
where
\beq
\o_r &=& e^{\zeta_r} = e^{\xi_r+ i\eta_r}, ~~~\op_s = e^{\xi^\prime_r + i\eta^{\prime}_s}, \\
(\o_+,\o_-) &=&  \left\{ 
\begin{array}{ll}
(\o_r, \op_s) , & ~~\mbox{for} ~~\xi_r \ge \xi^\prime_s   \\ 
(\op_s, \o_r) , & ~~\mbox{for} ~~\xi_r \le \xi^\prime_s 
\end{array} \right. 
\eeq

\subsection{Integral Formulas for $\bar N^{rs}_{nm}$}

\begin{itemize}
\item $\bar N^{rs}_{00}$ for $r \not=s$: \\
Taking $\zp \rightarrow Z_s$, ($\zeta^\prime_s \rightarrow -\infty$), 
\beq
N(\r_r, \rp_s) = 2 \sum_{n\ge 0} \bar N^{rs}_{n0} e^{n \xi_r} \cos (n\eta_r) = 2 \ln \vert z - Z_s \vert .
\eeq
Taking the limit where $z \rightarrow Z_r~~(\xi_r \rightarrow -\infty)$
\beq
N(\r_r, \rp_s) = 2 \bar N^{rs}_{00} =2 \ln \vert Z_r - Z_s \vert .
\eeq
Thus, 
\beq
\bar N^{rs}_{00} =\vert Z_r- Z_s\vert, ~~~\text{for}~~ r\not=s . 
\eeq

\item $\bar N^{rs}_{00}$, for $r = s$, \\
If we take the limit $z^\prime_r \rightarrow Z_r$ of Eq. (\ref{neumanna1}), 
\beq \label{neumanna8}
N(\rho_r, \rho^\prime_s) &=& 2\xi_r + 2\sum_n \bar N^{rs}_{n0} e^{|n| \xi_r} \cos \eta_r = 2\ln \vert z-Z_r \vert .
\eeq 
Taking the limit $z_r \rightarrow Z_r$ of Eq. (\ref{neumanna8}) again, we find 
\beq 
\bar N^{rr}_{00} &=& \ln |z_r-Z_r| - \xi_r. 
\eeq 
Since near $z_r = Z_r$, we expand $e^{-\z_r}$ 
\beq \label{c0058}
e^{-\z_r} &=& \frac{a_r}{(z_r-Z_r)} + \sum_{n=0} c^{(r)}_n (z_r-Z_r)^n .
\eeq 
It follows that in the limit $z_r \rightarrow Z_r$
\beq \label{limit61}
\xi_r  = \ln |z_r - Z_r| - \ln a_r .
\eeq
Using  Eq. (\ref{c0058})  and Eq. (\ref{limit61}) in the limit, where $z_r \rightarrow Z_r$, we obtain
\beq\label{Nrr00}
\bar N^{rr}_{00} &=& \ln a_r . 
\eeq 

\item $\bar N^{rs}_{n0}$, $n\ge 1$: Differentiating Eq.~(\ref{Nexpan2}) with respect to $\zeta_r$ (assume
$\xi_r \ge \xi^\prime_s$), 
\beq
\frac{\p }{\p \zeta_r} N(\r_r, \rp_s) &=& \frac{1}{2} \left(\frac{\p z}{\p \zeta_r} \right)
\left(\frac{1}{z-\zp} + \frac{1}{z-z^{\prime *}} \right) \nn\\
&=& \d_{rs} \left\{\frac{1}{2} \sum_{n \ge 1} \o^{-n}_r (\o^{\prime n}_s +\o^{\prime * n}_s ) +1 
\right\} \nn\\
&&+ \frac{1}{2} \sum_{n,m \ge 0} n \bar N^{rs}_{nm} \o^n_r(\o^{\prime m}_s + \o^{\prime * m}_s) \label{Diff1}
\eeq
where we use
\beq
\frac{\p}{\p \zeta_r} = \o_r \frac{\p}{\p \o_r} = \frac{1}{2} \left(\frac{\p}{\p \xi_r} -i \frac{\p}{\p \eta_r} \right) .
\eeq
Taking the limit, $\zp \rightarrow Z_s$ $(\o_s \rightarrow 0)$, 
\beq
\d_{rs} + \sum_{n \ge 1} n \bar N^{rs}_{n0} \o^n_r = \left(\frac{\p z}{\p \zeta_r} \right)
\frac{1}{z-Z_s}  
\eeq
By evaluating the contour integral around $\o_r = 0~ (z = Z_r)$, 
\beq
\oint_{\o_r = 0} d\o_r\, \o^{-n-1}_r \left(\d_{rs} + \sum_{m \ge 1} \bar N^{rs}_{m0} \o^m_r \right) = 
\oint_{\o_r = 0} d\o_r\, \o^{-n-1}_r  \left(\frac{\p z}{\p \zeta_r} \right) \frac{1}{z-Z_s} 
\eeq
we obtain 
\beq
\bar N^{rs}_{n0} &=& \bar N^{sr}_{0n} = \frac{1}{n} \oint_{Z_r} \frac{d z}{2\pi i} \frac{1}{z-Z_s} e^{-n\zeta_r(z)}, ~~~n \ge 1. \label{Nn0}
\eeq
Here we use $d\o_r = \o_r d \zeta_r$ and
\beq
\frac{d \o_r}{\o_r} \frac{\p z}{\p \z_r} = d \z_r \frac{\p z}{\p \z_r} = dz .
\eeq

\item  $\bar N^{rs}_{0n}$, $n\ge 1$: Differentiating Eq.~(\ref{Nexpan2}) with respect to $\zeta_s^\prime$ (assume $\xi^\prime_s \ge \xi_r$), 
\beq
\frac{\p }{\p \zeta_s^\prime} N(\r_r, \rp_s) &=& -\frac{1}{2} \left(\frac{\p z^\prime}{\p \zeta_s^\prime} \right)
\left(\frac{1}{z-\zp} + \frac{1}{z^*-z^{\prime}} \right) \nn\\
&=& \d_{rs} \left\{\frac{1}{2} \sum_{n \ge 1} \o^{\prime-n}_s (\o^{ n}_r +\o^{ * n}_r ) +1 
\right\} \nn\\
&&+ \frac{1}{2} \sum_{n,m \ge 0} m \bar N^{rs}_{nm} \o^{\prime m}_s(\o^{n}_r + \o^{ * n}_r) \label{Diffs}
\eeq
Here we make use of 
\beq
\frac{\p}{\p \zeta^\prime_s} = \o_s^\prime \frac{\p}{\p \o_s^\prime} = \frac{1}{2} \left(\frac{\p}{\p \xi_s^\prime} -i \frac{\p}{\p \eta_s^\prime} \right) .
\eeq
Taking the limit $z \rightarrow Z_r$ $(\o_r \rightarrow 0)$, 
\beq
\d_{rs} + \sum_{m \ge 1} m \bar N^{rs}_{0m} \o^{\prime m}_s = \left(\frac{\p z^\prime}{\p \zeta_s^\prime} \right)
\frac{1}{z^\prime-Z_r}  
\eeq

By evaluating the contour integral around $\o^\prime_s = 0~ (z^\prime = Z_s)$, 
\beq
\oint_{\o_s^\prime = 0} d\o_s^\prime\, \o^{\prime-m-1}_s \left(\d_{rs} + \sum_{m \ge 1} m\bar N^{rs}_{0m} \o^{\prime m}_s \right) = 
\oint_{\o_s^\prime = 0} d\o_s^\prime\, \o^{\prime-n-1}_s  \left(\frac{\p z^\prime}{\p \zeta_s^\prime} \right) \frac{1}{z^\prime-Z_r} 
\eeq
we obtain 
\beq
\bar N^{rs}_{0m} &=& \frac{1}{m} \oint_{Z_s} \frac{d z^\prime}{2\pi i} \frac{1}{z^\prime-Z_r} e^{-m\zeta^\prime_s(z^\prime)}, ~~~m \ge 1. \label{N0m}
\eeq
Here we use $d\o_s^\prime = \o_s^\prime d \zeta_s^\prime$ and
\beq
\frac{d \o^\prime_s}{\o^\prime_s} \frac{\p z^\prime}{\p \z^\prime_s} = d \z_s^\prime \frac{\p z^\prime}{\p \z^\prime_s} = dz^\prime .
\eeq

\item $\bar N^{rs}_{nm}$ : Differentiating Eq.~(\ref{Diff1}) with respect to $\zeta^\prime_s$
\beq \label{Diff2}
\frac{\p }{\p \zeta_r}\frac{\p }{\p \zetap_s} N(\r_r, \rp_s) &=& 
\frac{1}{2} \left(\frac{\p z}{\p \zeta_r} \right) \left(\frac{\p \zp}{\p \zetap_s}\right)
\frac{\p}{\p \zp} \left(\frac{1}{z-\zp} + \frac{1}{z-z^{\prime *}} \right) \nn\\
&=& \frac{1}{2} \left(\frac{\p z}{\p \zeta_r} \right) \left(\frac{\p \zp}{\p \zetap_s}\right)
\frac{1}{(z-\zp)^2} \nn\\
&=& \d_{rs} \frac{1}{2} \sum_{n \ge 1} n \o^{-n}_r \o^{\prime n}_s + \frac{1}{2} \sum_{n, m \ge 1} nm
\bar N^{rs}_{nm} \o^n_r \o^{\prime m}_s .
\eeq
Then evaluating the contour integral around $\op_s = 0$ $(\zp = Z_s)$, 
$\oint d\o_r \oint d\op_s \o^{-n-1}_r \o^{\prime -m -1}_s$ of Eq.~(\ref{Diff2}), 
we obtain 
\beq
\bar N^{rs}_{nm} &=& \frac{1}{nm} \oint_{Z_r} \frac{dz}{2\pi i} \oint_{Z_s} \frac{d \zp}{2\pi i} 
\frac{1}{(z_r-\zp_s)^2} e^{-n\zeta_r(z) - m \zeta^\prime_s(\zp)}, ~~~ n, m \ge 1  \label{Nnm}
\eeq 

\end{itemize}

\end{appendix}


%

%


\end{document}